# A BIBLIOMETRIC ANALYSIS ON SPECTRUM SENSING IN WIRELESS NETWORKS

[1]Nyashadzashe Tamuka[*]

University of Fort Hare, P.O Box X1314
Alice,5700, South Africa

201516429@ufh.ac.za

[2]Khulumani Sibanda

Walter Sisulu University, Eastern Cape
South Africa
ksibanda@wsu.ac.za

**Abstract**

**Spectrum scarcity is a prevalent problem in wireless networks due to network regulatory bodies' strict allotment of the spectrum (frequency bands) to licensed users. Such an operation implies that the unlicensed users (secondary wireless spectrum users) have to evacuate the spectrum when the primary wireless spectrum users (licensed users) are utilizing the frequency bands to avoid interference. Cognitive radio alleviates the spectrum shortage by detecting unoccupied frequency bands. This reduces the underutilization of frequency bands in wireless networks. There have been numerous related studies on spectrum sensing, however, few studies have conducted a bibliometric analysis on this subject. This study's goal was to conduct a bibliometric analysis on the optimization of spectrum sensing. The PRISMA methodology was the basis for the bibliometric analysis to identify the limitations of the existing spectrum-sensing techniques. The findings revealed that various machine-learning or hybrid models outperformed the traditional techniques such as matched filter and energy detectors at the lowest signal-to-noise ratio (SNR). SNR is the ratio of the desired signal's magnitude to the background noise magnitude. This study, therefore, recommends researchers propose alternative techniques to optimize (improve) spectrum sensing in wireless networks. More work should be done to develop models that optimize spectrum sensing at low SNR.**

*Keywords*: **spectrum sensing, bibliometric analysis, PRISMA, bibliometrics, R, signal-to-noise-ratio (SNR), spectrum, optimization.**

## 1. Introduction

The Covid-19 pandemic abruptly increased the demand for spectrum across the globe due to many organizations working remotely. South Africa's Independent Communications Authority (ICASA) temporarily released the 700MHz, 800MHz, 2300MHz, 2600MHz, and 3500MHz bands until November 2020 [Gillwald, 2020]. This was implemented to reduce network congestion and allowed operators to use 5G networks while providing low-cost services [Gillwald, 2020]. Thus, various governments generated revenue and created employment by licensing telecommunications companies to operate in specific wireless frequency bands. Owing to the global exponential growth of wireless networks, the demand for frequency bands has drastically increased across the globe. The regulatory bodies manage the spectrum by granting each operator (primary user) a license to operate within a specific frequency band. This means secondary (unlicensed) users must vacate the spectrum when primary (licensed) users are utilizing the band to avoid interference. Nevertheless, the spectrum management by these regulatory bodies tends to be rigid since most of the spectrum remains underutilized [Supraja and Pitchai, 2019]. Efficient and effective techniques should be in place to curb spectrum scarcity.

## 2. Background

### 2.1. *Cognitive Radio*

This section provides an overview of wireless cognitive radio technology. Its functionality is discussed, with a focus on spectrum sensing. A wireless cognitive radio is described as "a radio or system that senses its operational electromagnetic environment and can dynamically and autonomously adjust its radio operating parameters to modify system operations such as maximize throughput, mitigate interference, facilitate interoperability, and access secondary markets"[Federal Communications Commission, 2005]. Algorithms can be employed for sensing the wireless spectrum. Primary users (licensed users), as well as secondary users (unlicensed users), are the two types of spectrum consumers [Nasser et al., 2021]. The primary (licensed) users have priority in using the spectrum. Unlicensed users, classified as secondary users have lower spectrum occupancy priority and use the spectrum opportunistically. Spectrum distribution, spectrum movement, spectrum management, and also spectrum detection/sensing are all functions of cognitive radios.

The essential functions of the cognitive radio consist of (i) **observing** (tracking the accessible frequency bands and collecting vacant spectrum information ), (ii) **orienting** (determining the collected spectrum signal details by recognizing functional connections between system architectures and measurements), (iii) **learning** (evaluating the orientation phase's outcome by acquiring information to be applied in the future to boost decision capability), (iv) **deciding** (choosing necessary frequency bands according to their characteristics as well as user information), and v) **acting** (carrying out actions by making effective use of available bands). Fig 1 depicts a set of activities known as the cognitive. Fig 1 depicts a set of activities known as the cognitive cycle [Jain et al., 2019].

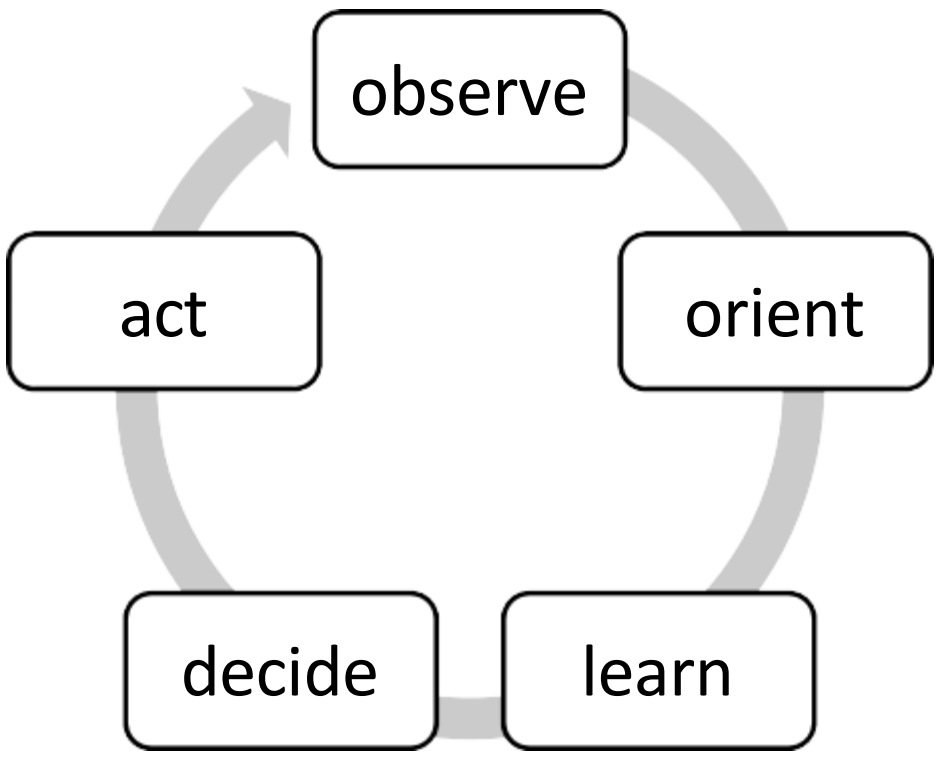

Fig 1: The cognitive cycle

### 2.2. *Spectrum sensing*

Spectrum sensing (spectrum detection) is a fundamental operation carried out by cognitive radio technologies. It allows unlicensed users to gain insight into the radio environment by identifying the existence of licensed user signals and determining whether to transmit within its frequency band [Arjoune and Kaabouch, 2019]. The spectrum detection model is written as follows:

$$y(n) = \begin{cases} v(n) & \text{H}_0\text{: PU absent} \\ h * s(n) + v(n), & \text{H}_1\text{: PU present} \end{cases} \quad (1)$$

From Eq. (1), $\boldsymbol{n}$ represents the sample number ranging from 1 to N, where N is the total sample amount. The signal acquired by the unlicensed user is denoted as $\boldsymbol{y(n)}$, the signal propagated by the licensed user is $\boldsymbol{s(n)}$, and $\boldsymbol{v(n)}$ is the noise signal. The channel amplitude gain of the sensing channel, such as an antenna, is indicated as $\boldsymbol{h}$. The primary user's absence and presence are denoted as $\boldsymbol{H_0}$ and $\boldsymbol{H_1}$ respectively. From Eq. (2) the decision on the primary user signal's presence is established by comparing the detector's output, which is also called the test statistic, to the threshold. The decision is computed as:

$$\begin{cases} \text{if } S \geq \gamma, & H_1 \\ \text{if } S < \gamma, & H_0 \end{cases} \quad (2)$$

From Eq. (2), $S$ represents the test statistic and $\boldsymbol{\gamma}$ denotes the threshold. The detector's performance is assessed using the "*Probability of detection*/$\boldsymbol{P_d}$" and the "*Probability of false alarm*/ $\boldsymbol{Pfa}$"

$\boldsymbol{P_d}$ in Eq. (3) is the likelihood that $\boldsymbol{S}$ correctly determines $\boldsymbol{H_1}$

$$\boldsymbol{P_d} = \boldsymbol{P}\{\text{ decision } = \boldsymbol{H_1}/\boldsymbol{H_1}\} = \boldsymbol{P}\{\boldsymbol{S} > \boldsymbol{\gamma}\ /\boldsymbol{H_1}\} \quad (3)$$

$\boldsymbol{Pfa}$ in Eq. (4) is the likelihood that S determines $\boldsymbol{H_1}$ while it is $\boldsymbol{H_0}$

$$\boldsymbol{Pfa} = \boldsymbol{P}\{\text{ decision } = \boldsymbol{H_1}/\boldsymbol{H_0}\} = \boldsymbol{P}\{\boldsymbol{S} > \boldsymbol{\gamma}\ /\boldsymbol{H_0}\} \quad (4)$$

Traditional algorithms have been proposed for sensing spectrum such as wavelet-based, matched filter, cyclostationary feature-based detection, and energy detector [Arjoune and Kaabouch, 2019].

### *3.1.1. Energy detection*

This approach approximates the signal's power relative to a predetermined threshold, determining the absence or presence of the licensed user [Arjoune and Kaabouch, 2019]. This technique aims to determine between two hypotheses and Energy detection can be presented as Eq. (5):

$$\begin{aligned} &y(t) = x(t), H_0 && \text{(Primary User absent)} \\ &y(t) = as(t) + x(t), H_1 && \text{(Primary User present)} \end{aligned} \quad (5)$$

$y(t)$ is the input signal, $s(t)$ is the signal propagated by the licensed user, $x(t)$ is the noise signal added to the received(input) signal, and $a$ is the channel's gain in amplitude. The null hypothesis, $H_0$ asserts that the frequency band is unoccupied (the licensed user is absent). Due to its minimal computational and technical complexities, energy detection is an extensively applied method for spectrum sensing [Shah and Yelalwar, 2019]. It is a detection technique for sensing the licensed user signal using the FFT (Fast Fourier transform) [Shah and Yelalwar, 2019]. The FFT converts digital signals from a time dimension to a frequency-domain form and generates the power for all frequencies, resulting in a PSD / power spectrum density. Fig 2 presents the energy detection process.

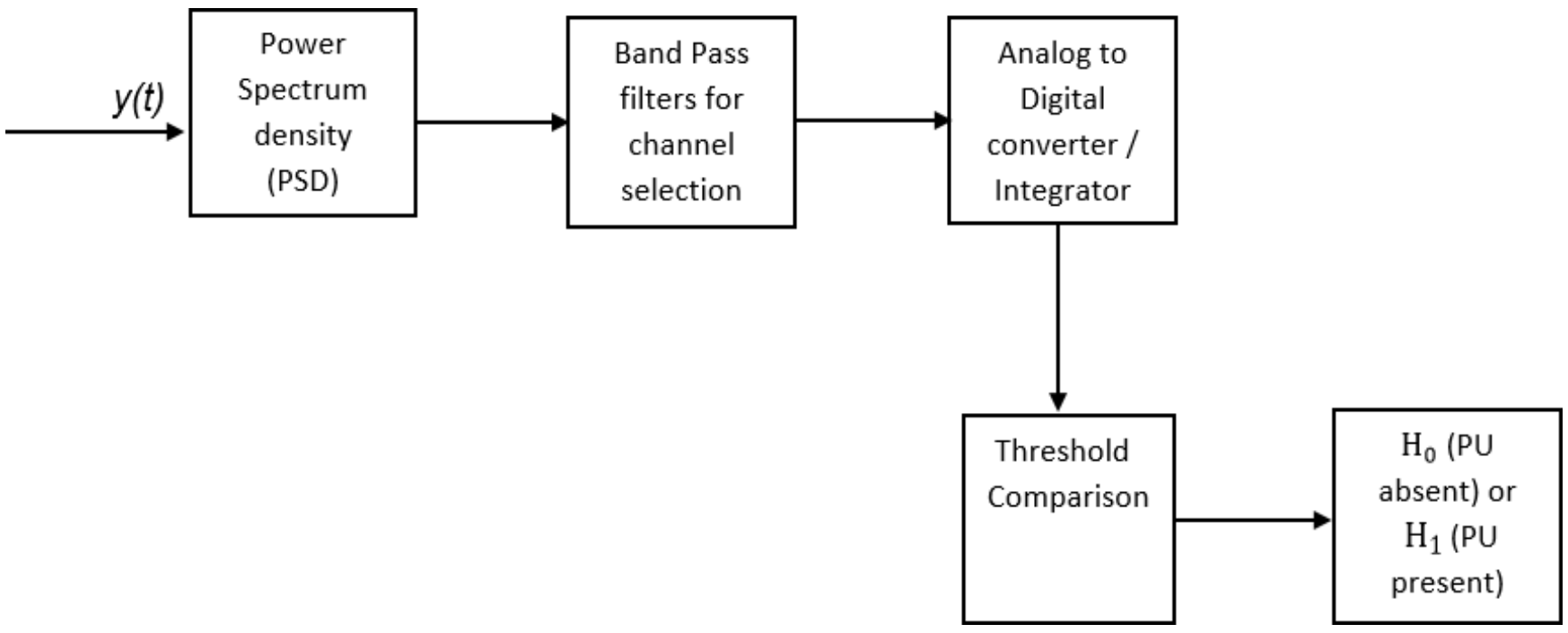

Fig 2: The energy detection process.

The input signal $\boldsymbol{y(t)}$ is applied to a bandpass filter (BPF) to choose a channel and is transformed by an Analog to Digital (A/D) transformer over some time [Arjoune and Kaabouch, 2019]. The output is subsequently compared to a threshold predefined for determining whether the PU is present. Depending on the channel's state, the threshold value can be set to be either constant or variable [Arjoune and Kaabouch, 2019].

### *3.1.2. Matched filter*

This technique necessitates a detailed prior understanding of the signal information by the primary user. This information which consists of the operating frequency, bandwidth, pulse shaping and modulation is applied for demodulating the signal received [Salahdine, 2017]. This technique is thus regarded as efficient for spectrum detection because it necessitates the least processing time to compute the $Pfa$ [Salahdine, 2017]. However, the drawback of this technique is that it calls for prior information from the licensed user, thus consuming more power and high computational complexity. Fig 3 presents the matched filtering process.

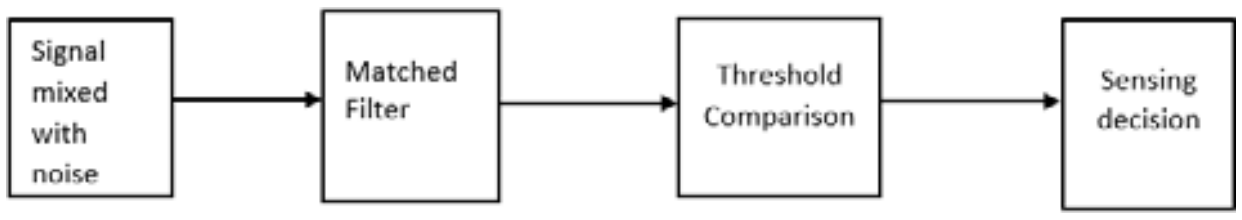

Fig 3: the matched filter detection.

The technique is presented by Eq. (6):

$$Y(n) = \sum_{a=\infty}^{\infty} h(n-a)x(a) \quad (6)$$

Where $Y(n)$ is the unlicensed used received signal, h is the impulse response of a matched filter detector that is matched to the reference signal n, and x is the unknown signal convolved to maximize SNR. By contrasting the output of the matched filter with the predetermined threshold and the signal's time-shifted version, the matched filter may detect the presence of the licensed user. This technique is beneficial only if the cognitive network is aware of the primary user signal.

*3.1.3. Wavelet-based spectrum detection*

Spectrum sensing is essential in cognitive radio technology for efficient bandwidth utilization. Interference between adjacent frequency bands can be efficiently avoided by carefully determining spectrum boundaries. Wavelet edge detection is a technique for detecting spectral boundaries accurately [Monga et al., 2022]. Fig 4 presents the wavelet-based spectrum detection process.

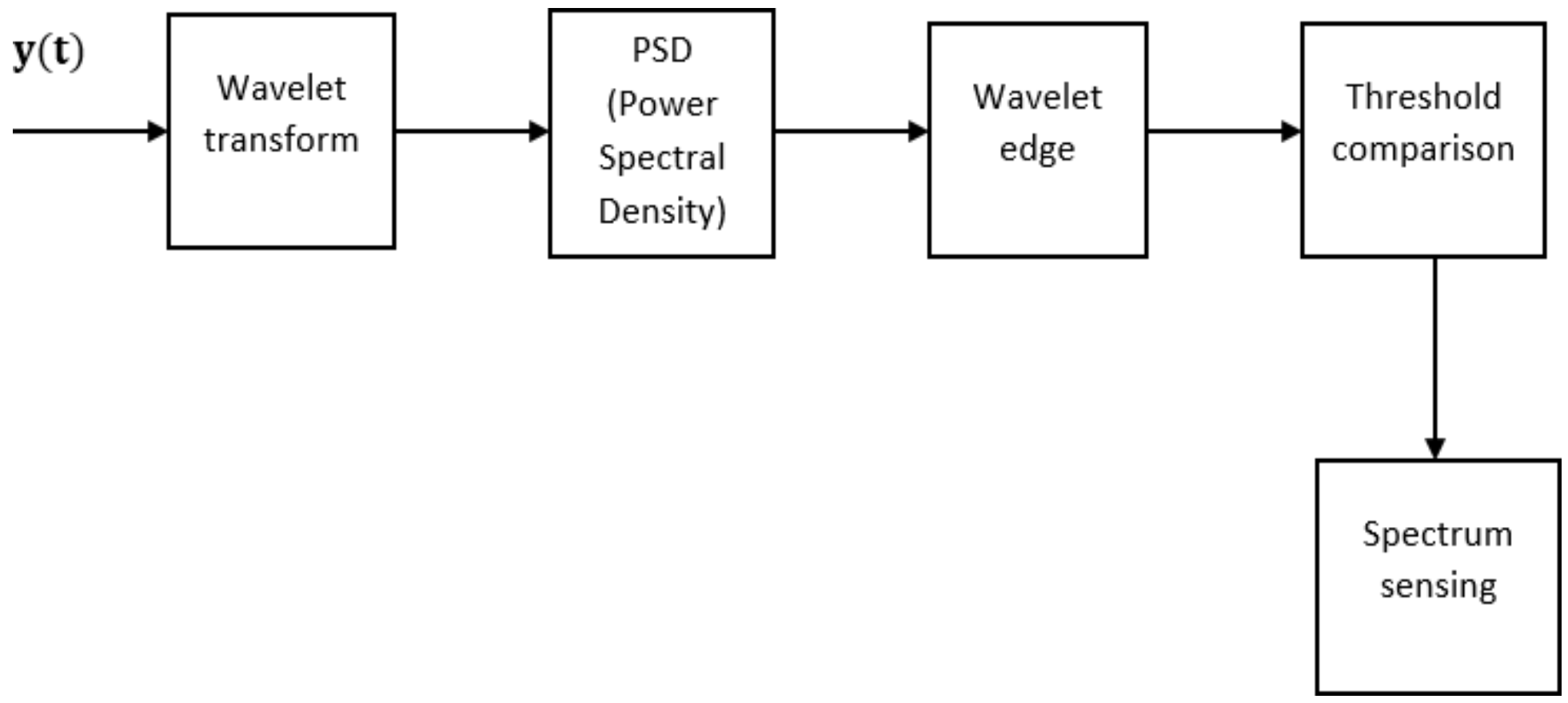

Fig 4: wavelet-based spectrum detection

Although this method is widely used, the system's efficiency is quite low depending on the type of wavelet used. A spectrum sensing technique selects the best wavelet function for the supplied spectrum after analyzing the nature of peaks in the frequency bands' power spectra. A wavelet is a waveform of limited duration with amplitude starting at 0 and returning to 0 after oscillation [Monga et al., 2022]. The continuous wavelet transform, which enables the detection of signal-decomposed coefficients using a basis [Monga et al., 2022], serves as the foundation for wavelet sensing and edge detection. The continuous wavelet function, $\psi(t)$ is applied for a given signal, $x(t)$ which includes zeros outside of a specific range and works in the time domain [Monga et al., 2022]. It is presented by Eq. (7):

$$f(c,w) = \langle x(t), \psi_{u,w} >= \int_{-\infty}^{+\infty} x(t)\psi_{u,w}^{*}(t)dt \quad (7)$$

Where $w$ is the scaling parameter, $c$ is the translating parameter and $\psi_{u,w}$ is the basis. The wavelet-based sensing calculates the power spectral density using the continuous wavelet transform of the signal. A one-dimensional signal, $x(t)$ is transformed into two-dimensional coefficients, $f(c, w)$ by the wavelet transform. When the frequency matches the parameter $c$ and the time instant matches the parameter, the frequency-time analysis can be performed. The sensing result is presented in Eq. (8):

$$\begin{cases} \text{If } e \geq \lambda, & \text{PU not present} \\ \text{If } e < \lambda, & \text{PU present} \end{cases} \quad (8)$$

where the sensing threshold is $\lambda$ and $e$ is the wavelet edge. Wavelet-based sensing has been presented to detect the proportion of occupied bands throughout a frequency range, although it takes a long time to analyze [Monga et al., 2022].

### *3.1.4. Cyclostationary detection*

This approach distinguishes the modulated signal from noise [Arjoune and Kaabouch, 2019]. A signal is considered to be cyclostationary if its autocorrelation and mean are periodic functions [Verma et al., 2012]. The method of obtaining features from an input signal and carrying out detection based on the extracted features is referred to as feature detection. Cyclostationary detection can distinguish primary user signals from noise and can be applied at low SNR by utilizing the primary user signal's information that is not present within the noise signal. The main disadvantage of this method is its calculation complexity. Furthermore, it has to deal with every frequency to construct the spectral correlation function, which results in a complex calculation. The benefit of feature detection above energy detection is that it allows for differentiating between various signals or waveforms [Verma et al., 2012]. Fig 5 indicates the cyclostationary detection technique process.

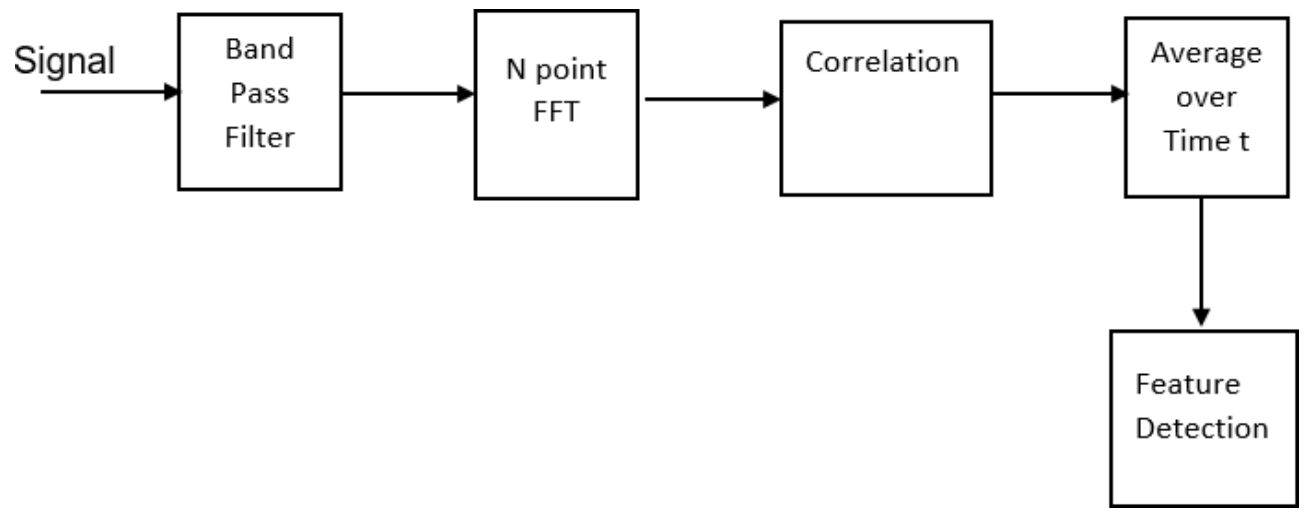

Fig 5: the cyclostationary detection process

For signal detection, cyclostationary-based sensing employs the cyclic-correlation expression, widely referred to as the spectral correlation density, as opposed to the Power Spectral Density (PSD) used by the energy detector. Cyclostationary feature detector uses the cyclicity from the user signal acquired for determining the primary users' (PU) presence. The spectral correlation density is computed by Eq. (9):

$$S(f,\alpha) = \sum_{\tau=-\infty}^{\infty} R_y^{\alpha}(\tau) e^{-j2\pi f\tau} \quad (9)$$

where $\boldsymbol{R_y^{\alpha}(\tau)}$ denotes the cyclic auto-correlation-based function computed by the expression:

$$\boldsymbol{R_y^{\alpha}(\tau) = E[y(n+\tau)y^{*}(n-\tau)e^{-j2\pi\alpha n}]} \quad (10)$$

where $\boldsymbol{\alpha}$ is the cyclic frequency.

There have been few bibliometric kinds of research on the optimization of spectrum sensing. Bibliometrics is defined by [Ebrahim, 2017] as "the statistical analysis of publications with a focus on quantitative analysis of citations". [Aria and Cuccurullo, 2017] asserted that bibliometrics can implement a transparent, systematic, and duplicable review utilizing the scientific activity's statistical measurement. The research output in spectrum detection was analyzed in this study. The study aimed to conduct a bibliometric analysis on the optimization of spectrum sensing (SS) in wireless networks. The study addresses two research questions, which are: What are the existing traditional spectrum sensing techniques in cognitive radio networks? To what extent are machine-learning algorithms applied to optimize spectrum sensing in wireless networks?

### **2.3. *Overview of studies on spectrum sensing***

[Ivanov et al., 2021] presented a comprehensive study of stochastic spectrum sensing techniques categorized by the attributes they derive from input signal samples to optimize spectrum sensing in 6G networks. Their objective was to accurately detect the licensed user signal fused with noise signals. The key design characteristics included $P_d$, computational complexity, fading and resistance to noise, pros and cons as well as signal-to-noise model assumptions. Their findings revealed that in contrast to conventional methods, deep learning methods are widely proposed for spectrum sensing in 6G networks. [Shah and Koo, 2018] supported [Ivanov et al., 2021] by establishing a reliable spectrum sensing scheme centered on the K-nearest neighbor. The spectrum dataset was simulated in Matlab. The authors demonstrated that throughout the training stage, all CR/cognitive radio users provided a sensing response under varied parameters and either transmitted or remained silent based on a global decision. Sensing reports were organized into classes, and the report was assigned to the classes. The PU was declared absent or present based on the classification result. They projected that a CR user was transmitted or remained silent based on the global decision that was contrasted to the primary user activity, as determined signal acknowledged by each CR user. They evaluated performance using metrics such as the total probabilities of error and detection, as well as the capacity to take advantage of data transmission opportunities. Their simulation findings presented that their proposed scheme outperformed the traditional schemes in both nonfading and fading environments. Nevertheless, the weakness of their study is that it does not demonstrate spectrum detection without prior primary user knowledge.

A study by [Sarala et al., 2020] highlighted that at uncertain noise levels, the licensed users resisted interference, and the proposed ATSED significantly improved the spectrum detection performance. The information signal was subjected to band-pass filtering using an advanced dynamic threshold technique to analyze the power range of the licensed user for their energy signal. Matlab was applied for establishing a spectrum detection framework. The Pfa, as well as the $P_d$, were for evaluating the model's performance. The simulation results showed that the signal-noise-ratio (SNR) increased, and so did the $P_d$. It was also observed that as the bandwidth factor increased, the detection of probability decreased. Again, the drawback of this study is that it does not demonstrate spectrum detection without prior primary user knowledge.

In a pursuit to optimize spectrum detection at the lowest signal-noise-ratios (SNR), [Mahendru et al., 2020] proposed a novel mathematical algorithm for determining the best sensing period/number of samples) for an energy detection method in the presence of noise uncertainty. The Pfa, as well as the $P_d$, were adopted as evaluation metrics. For generating a list of SNR values, they varied the number of samples from 100 to 20000. The proposed model was developed through simulations in Matlab. The effect of noise on the number of samples sensed was investigated, and a proposed novel approach for correlating the sensing period with SNR to yield a desired model performance was presented. The authors argued that their approach had the potential to be useful for spectrum detection in low SNR scenarios for cognitive radio networks.

The study by [George and Prema, 2019] pinpointed that by using the frequency/periodicity amongst the modulated signals' autocorrelation, cyclostationary feature detection (CFD) can be utilized for detecting primary users (PU). The authors revealed that since the noise was uncorrelated, these algorithms attempt to distinguish between signal and noise. The drawback of CFD is that detection requires prior knowledge of the PU signal (semi-blind). Based on CFD, [George and Prema, 2019] proposed a novel algorithm for completely blind detection performance. The simulations to develop a model were conducted in Matlab. The statistical attributes of the cyclic spectrum were used to classify PU signals. Furthermore, an algorithm was developed to identify the signal's modulating method to detect and classify it without the use of any training algorithms. The proposed method was 100% accurate to detect PU reliably at notably low SNRs such as -7 dB and -5 dB, without prior knowledge of PU or noise in the channel. The probability of misdetection (false alarm) was constant at approximately 0.1 for all SNR values. Although the proposed model was accurate in spectrum detection, computational complexity is its drawback.

Even at low SNR levels, spectrum sensing centered around the cyclostationary-feature algorithm can be applied to accurately detect the presence of licensed users. However, detecting modulated signals at significantly low SNR levels necessitates a greater amount of analyzed samples [Lima et al., 2018]. In their study, [Lima et al., 2018] proposed a spectrum sensing architecture that reduced the execution time required for determining the cyclostationary features' signal when multi-core processors were employed. The proposed architecture's performance was evaluated using parallel speedup and parallel efficiency as metrics. Their findings revealed that the proposed architecture significantly reduced the execution time. The authors hypothesized that the time saved as a result of parallel computing can be applied for enhancing the throughput and reducing the spectrum collision period. Despite this, it can be noted that the whole proposed spectrum detection method is a cumbersome process.

[Odhavjibhai and Rana, 2017] contributed to the spectrum detection (sensing) notion by proposing a matched filter approach based on various parameters. The spectrum detector was created using simulation methodology. The simulations were carried out in MATLAB, as with other spectrum sensing studies in the literature. The possibility of a false alarm and the detection probability was utilized for evaluating the performance of the matched filter detector. The findings showed that as the SNR increased, so did the probability of detection. Furthermore, as the number of samples increased, so did the probability of detection and the SNR. Although the proposed model performed well for low SNR, its limitation is that it requires prior primary user information.

The study by [Odhavjibhai and Rana, 2017] was alluded to by [Brito et al., 2021] who argued that a hybrid matched filter detection (HMFD) can outperform other conventional sensing techniques. The authors carried out the simulations in MATLAB. The simulations sought to understand how parameters such as false alarm probability, SNR, and sample number affect the probability of miss-detection. The results revealed that in terms of detecting the PU's presence, the HFMD/hybrid matched-filter detector outperformed the conventional methods using a low signal-to-noise ratio, a small number of samples, and a false alarm probability slightly less than 0.5. The weakness of this study is that it does not take into account scenarios in which prior primary user information is unknown.

To address the issue of spectrum scarcity, [Moawad et al., 2020] suggested a wideband spectrum detection method based on cepstral analysis. To begin, they proposed an adaptive log-spectral density technique for the edge determination phase, detecting spectral limits. They also presented the mathematical framework for the suggested algorithm and obtained the formula for determining the sensing threshold of the recommended detector. At low-to-medium-noise decibel levels, the simulation results demonstrated that the edge identification algorithm beat different wavelet-based algorithms. The proposed edge detector performed well at low signal-to-noise ratios when used in conjunction with denoising. They introduced an improved passband auto spectrum detector for the primary user detection phase to address the misdetection problem, and it outperformed numerous approaches. Notwithstanding the high level of performance yielded by the suggested wavelet-based approach, its computational complexity is a remarkable drawback.

The study by [Negi et al., 2019] hypothesized that wavelet entropy detection was noise-independent. The authors proposed a scheme in which the received signal was efficiently decomposed at the receiver to improve entropy results. The study recommended deriving Wavelet components from the received signal instead of the traditional method of deriving DFT / discrete Fourier transform components. The wavelet entropy was subsequently compared to the threshold value to establish whether or not the Primary User (PU) was using the frequency band. The wavelet packet transforms were utilized in their study to decompose the received signal before calculating entropy, which was then used to identify PU. The wavelet-based detector's performance was assessed using the ROC curves, false alarm probability, and detection probability. Their findings revealed that the wavelet entropy-based detection performed accurately at low SNR levels for example -25 dB. Notwithstanding the hypothesis being true, this approach has some limitations, for instance, prior primary user knowledge is required. Consequently, this detector uses complex algorithms hence akin to high computational complexity.

Contrary to the study by [Negi et al., 2019], [Muñoz et al., 2022] executed the simulation experiments on an SDR (software-defined radio) testbed. In quest of licensed / primary user emulation, the ensemble, K-nearest neighbors (KNN), and the support vector machine (SVM) were adopted. The findings revealed that the SVM outperformed other techniques utilized by the authors (5% more than other approaches). Despite this, the proposed approach has some limitations, for instance, they incorporated the energy detector before training the algorithms. The energy detector was susceptible to low SNR levels. Contrarily, an ideal detector must be robust at low SNR levels for blind sensing (detecting primary users without prior knowledge).

To identify the primary user (PU) presence, [Saber et al., 2019] proposed a low-power-consumption and cost-effective spectrum sensing operation using real signals. A 433 MHz transmitter and a Raspberry-pi device were adopted to generate the signals. An energy detector (ED), SVM, and artificial neural networks (ANN) were employed for detecting the signals. The study compared the performance of the classifiers using the ED/energy detector's output to determine the best method for spectrum sensing. The detection probability and as well as the false-alarm probability were applied for evaluating the proposed models' performance. The findings demonstrated that SVM was more accurate in spectrum sensing than its counterparts. The received signals were identified as PU or not (noise) using the ED approach, in addition to training and validation on proposed SVM and ANN detectors. Hence, the drawback of their study is that very low SNR magnitudes degrade models' performance. This is because the models' performance depends on the ED's output.

To curb the drawbacks of traditional sensing methods, [Ahmad, 2019] proposed a novel detection method based on machine learning. A novel ECD (cyclostationary feature-based signal detector and ensemble hybrid approach were proposed in their study. The ensemble classifier extracts cyclostationary features from communication signals and used them to detect the PU's signal. The authors simulated the dataset for their study using Algorithm 1 and included the noise configurations in a Gaussian medium with a minimum SNR. The signal-to-noise ratio varied between -5dB and -15 dB. The FFT accumulation method was adopted to extract cyclostationary features. The AdaBoost and decision trees were utilized for training the ensemble classifier. Furthermore, SVM was trained using the same extracted features for performance comparison. Classifiers were compared using a confusion matrix and ROC curves. The ensemble classifier was compared to another machine learning classifier, the support vector machine (SVM). The results revealed that the ensemble classifier outperformed the SVM. In addition, the simulation findings showed the superior efficiency and robustness of the detector when compared to a cyclostationary detector with no conventional energy and a machine learning detector. Even so, future research should concentrate on reducing the computational complexity of the cyclostationary feature-based signal detector.

On the other hand, the supervised ML algorithms, decision tree (DT), SVM and the k-nearest neighbor (kNN), were adopted by [Mohammad et al., 2022] for detecting the presence of licensed users across the TV band. Furthermore, a PCA/principle-component analysis was applied to catalyze the classifier's learning. Moreover, the ensemble approach was employed to improve the classifier's predictability and performance. The proposed model was assessed using real data collected from ten various locations throughout Windsor-Essex. The models' performances were evaluated and compared using the Receiver Operating Characteristic (ROC), F-measure, and accuracy. According to simulation results, the ensemble classifier achieved the best performance. Moreover, the simulation results indicated that using PCA shortened the training time while maintaining performance. However, the best performance came at the expense of increased computational complexity.

The SVM and KNN were employed by [Tamilselvi and Rajendran, 2023] to improve spectrum utilization. The detection probability was plotted using the KNN and SVM algorithms, with a constant false alarm probability. When secondary users were used, a ROC curve was plotted to inspect the spectrum. The performance of these two algorithms was compared based on a false-alarm rate, with the KNN algorithm outperforming the SVM. One of the study's drawbacks is that it does not highlight the performance of the models in the absence of prior licensed user knowledge (unsupervised dataset).

In a similar study, [Wasilewska and Bogucka, 2019] argued that machine learning algorithms can be applied for enhancing the reliability of spectrum sensing. Random Forest and k-Nearest Neighbours were proposed to enhance the detection performance. These recommended algorithms were implemented to Energy Vector-based/EV and Energy Detection/ ED input data for detecting the LTE/Long-Term-Evolution signal's presence for a 5G new radio system to utilize the available resource blocks. The findings revealed that for high SNR values, the proposed models (RF as well as kNN) based on EV as input data outperformed those based on energy-detection (ED) hard decisions. However, this study has several limitations for instance the proposed models require more computational capabilities and memory. In addition, the signal pre-processing approach succumbs to high noise signals (high false positive rate).

A similar study by [Valadão et al., 2021] proposed a collaborative spectrum detection system. The system was based on a hybrid random forest classifier (RF) and neural network. The experiment's algorithms were executed in Python version 3.7.9. The study aimed at developing a collaborative framework that could yield maximum accuracy at higher magnitudes of noise while involving fewer secondary users in the network. As a result, they proposed extracting features from each unlicensed user's sensing information, then using RF to classify the PU's presence in each frequency band. The data from various unlicensed users were subsequently shared with a fusion center, in which a residual neural network-trained model determined whether or not a licensed user was present. The model proposed performed well at high noise magnitudes, showing that the approach could determine the PU's presence in 98% of the situations when the assessed channel had a significant noise level. The findings were achieved through the collaboration of ten unlicensed users. Even at the lowest SNR, the authors yielded positive findings. Furthermore, the DLSenseNet exhibited a greater detection probability as compared to other approaches. Contrastingly, the study did not indicate how the proposed model could be applied to unsupervised data. Additionally, the proposed model incurred computational overhead.

A study by [Zheng et al., 2020] presented spectrum detection as a classification task. They proposed a deep learning classification-based sensing method. For compensating the consequences of noise, the authors normalized the received signal power. To enable the proposed ResNet technique for adapting to untrained signals, they trained it with as many various types of signals as possible mixed with noise signals. Transfer learning approaches were also adopted to boost performance for real spectrum signals. Extensive experiments were carried out to assess the performance of the technique. The simulation results revealed that the proposed approach was superior to the traditional approaches. However, the study used simulated signals and not real-world signals.

In an attempt to detect unoccupied frequency bands (spectrum), [Sarikhani and Keynia, 2020] proposed a Recurrent convolutional neural network (RNN)-based algorithm for cooperatively sensing spectrum holes. They used the convolutional layer to examine the spatial correlation as well as the LSTM layer to leverage the temporal correlation. As a result, the proposed scheme applied the spectrum spatiotemporal features to alleviate cooperative sensing and enhance sensing accuracy. A labeled dataset was generated for training the proposed network. They altered three main parameters stochastically for generating the labeled data. These three parameters were channel conditions, cooperating node positions, and the permutations of the collected spectrum data nodes in the proposed model's input. All simulations were carried out using MATLAB software. The probability of miss-detection, detection probability as well as false alarm probability was utilized at various SNR (signal-to-noise) ratios for evaluating the models' performance. The findings revealed that the LSTM model outperformed other proposed models in the literature. The limitation of the work by [Sarikhani and Keynia, 2020] is that they applied labeled data for simulation which does not present a real cognitive radio scenario.

Due to the rapid increase in demand for wireless technology, there is a high demand for wireless spectrum. A study by [Anand et al., 2020] adopted the KNN and SVM techniques to optimize spectrum sensing. They argued that the aforementioned techniques could be used to map an input based on examples into an output (from a set of predefined outputs). They used a dataset consisting of 1299 frequencies. The dataset was selected from GitHub. The dataset contained the frequencies of about 8 users. It included the signal with and without noise. An experiment was conducted at the constant SNR of -10db. The ROC missed detection as well as the probability of false alarms the metrics used in the simulation for comparing and evaluating the classifiers' performance. The simulation findings conveyed that at the considered false alarm probability, the KNN outperformed the SVM. In contrast, the drawback of the study was that it used a constant value of SNR (-10dB) and a supervised dataset.

Authors [Rajendran et al., 2018] investigated the modulation categorization for a scattered wireless network. The RadioML2016.10a dataset, which is an open-access dataset, was used for training and evaluating the classifier. Initially, a novel LSTM model for automatic modulation classification was established. Instead of complex attributes like higher-level periodic moments, the model learned from phase and amplitude information in the training dataset. Analyses presented that the proposed model achieved an accuracy of approximately 90% at several signal-to-noise ratios. Nevertheless, the proposed model has some drawbacks, for example, the LSTM model was tested on labeled datasets which is not the real world scenario.

## 3. Methods

A systematic literature review regarding spectrum-sensing (detection) was conducted for this study. The study addresses two research questions which are: What are the existing classical/traditional spectrum detection techniques in wireless networks? and to what extent are machine learning algorithms applied to optimize spectrum sensing in wireless networks? The Preferred-Reporting-Items for-Systematic-Reviews-and-Meta-Analyses (PRISMA) method [Selçuk, 2019] was adopted in this study. The PRISMA method is widely used for a systematic review that thoroughly scans all the articles published on the topic to discover answers to established research questions [Selçuk, 2019] and will do so by employing different criteria for exclusion and inclusion for identifying the relevant reports that the review consists of, and then summarize the findings [Selçuk, 2019] Using statistical methods to interpret the results necessitates a systematic review with meta-analysis [Selçuk, 2019]. The PRISMA method consists of four stages which are identification, screening, eligibility, and inclusion [Sarkis-Onofre et al., 2021]. Fig 6 shows the PRISMA's stages flow diagram.

### 3.1. *Identification of the articles*

Three thousand one hundred and fifty-one (3151) articles were identified from the WoS (Web of Science) and the Scopus databases. The data sources were selected based on their applicability to the domain of interest. Furthermore, the availability of multiple full-text and open-source articles was a motivator for selecting the databases in this review. This allowed easy accessibility of research articles and it eliminated bias [Edanz-Learning-Team, 2022]. Hence the aforementioned databases were adopted for retrieving the relevant articles. These databases were accessed online, which made it easy for obtaining the articles.

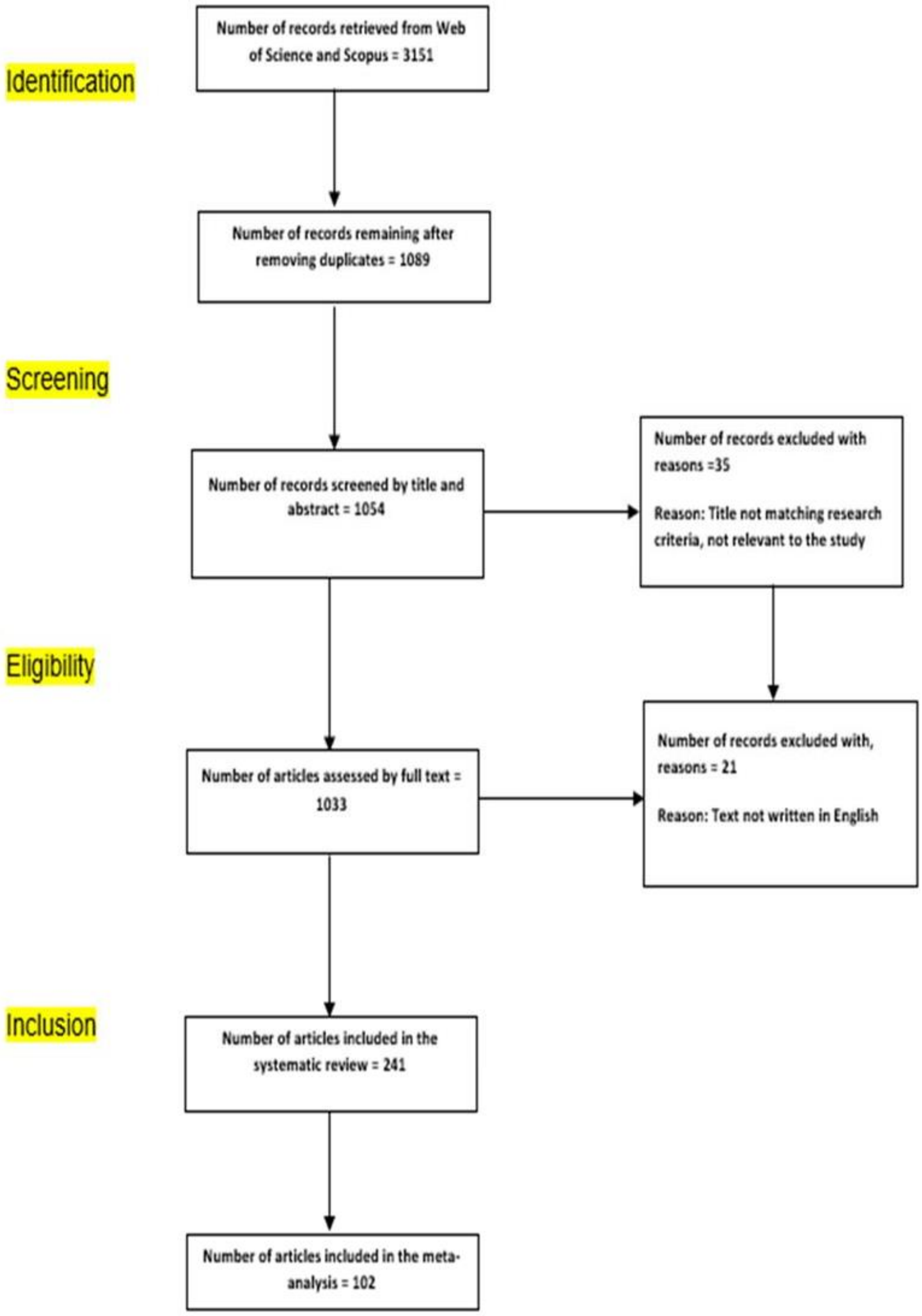

*Fig 6*: The PRISMA flow diagram [Zheng et al., 2020]

### 3.2. *Search Strategy*

The search was conducted on the selected databases using the following keywords: spectrum sensing and cognitive radio networks. The searches were performed by combining the keywords using the AND operator, provided the operator was allowed. The following search string was used to obtain the sources from the databases: spectrum sensing AND cognitive radio networks (Title) for the Web of Science as well as TITLE(spectrum AND sensing AND cognitive AND radio AND networks) for the Scopus database. One thousand three hundred and seventy (1370) documents were retrieved from the Web of Science and 1781 documents were retrieved from Scopus before the screening.

### 3.3. *Screening*

This involved reading the title and abstract of the identified articles. This was conducted to recognize documents that were relevant for systematic review. Articles that addressed the research questions were identified based on a yes or no decision. Some articles were excluded for a reason. The common reasons were that some articles were not open source and they were not relevant to the research questions and outcomes.

### 3.4. *Eligibility*

The initial stage was using the search strings to search the relevant articles for downloading. The restrictions (filtering criteria) were used to retrieve relevant documents. Only peer-reviewed journal and conference articles were considered for review. This was done to retrieve creditable documents for review. Articles from 2017 to 2022 were selected for ensuring that recent sources were utilized for review. Table 1 presents the search strategy and the inclusion criteria.

Table 1. Search strategy and the inclusion criteria

| Stage | Inclusion criteria | Documents retrieved from the Web of Science | Documents retrieved from Scopus |
|---|---|---|---|
| 1 | Choose all the documents that contain the words 'spectrum sensing' AND 'cognitive radio' in the title | 1370 | 1781 |
| 2 | Select the period 2017-2022 | 472 | 617 |
| 3 | Select only conference and journal articles | 467 | 587 |
| 4 | Select only documents written in English | 454 | 579 |
| 5 | Select only open-access full-text articles | 105 | 136 |

### 3.5. *Inclusion*

After filtering (exclusion and inclusion), the included articles were downloaded from the databases and exported as a BibTeX file. The data was imported and loaded into R data frames. These BibTeX files were then merged and the duplicates were removed. After removing duplicate records (records with the same title), and selecting relevant documents, 102 records were left for meta-analysis. Bibliometrix which is an R package was utilized for analysis [Derviş, 2019]. Bibliometrics is an open-source tool for conducting a thorough mapping study of literature [Derviş, 2019].

### 3.6. *Analysis*

*The sample*

The selected records were imported to Biblioshiny in xlsx format. Biblioshiny is a tool for bibliometric visualizations [Aria and Cuccurullo, 2017]. After applying the filtering criteria, 102 documents from 2017 to 2022 remained for analysis. These documents were written by 334 various authors and published in 49 different sources. The sample consisted of five different categories: 93 journal articles (91.18%), 1 book chapter article (0.98%), 2 early access articles (1.96%), and 6 proceedings papers (5.88%). This reveals that journal articles were a majority category. There were 332 authors for co-authored articles and 2 authors for manuscripts authored by one researcher. This reveals that the majority of the sample was multi-authored documents. In the same vein, regarding author collaboration, the number of single-authored documents was 2, the number of co-authors per document was 3.57 (approximately 4), and the international co-authorship proportion was 24.51%. Table 2 presents the sample.

Table 2. The sample

| **Description** | **Results** |
|---|---|
| Timespan | 2017-2022 |
| Sources (Journals, Books, etc) | 49 |
| Documents | 102 |
| Annual Growth Rate % | -6,51 |
| Document Average Age | 2,8 |
| Average citations per document | 7,814 |
| References | 1 |
| **DOCUMENT CONTENTS** | |
| Keywords Plus (ID) | 118 |
| Author's Keywords (DE) | 321 |
| **AUTHORS** | |
| Authors | 334 |
| Authors of single-authored documents | 2 |
| **AUTHORS COLLABORATION** | |
| Single-authored documents | 2 |
| Co-Authors per document | 3,57 |
| International co-authorships % | 24,51 |
| **DOCUMENT TYPES** | |
| article | 93 |
| article; book chapter | 1 |
| article; early access | 2 |
| proceedings paper | 6 |

The filtered data were analyzed using various bibliometrics features such as the word cloud, conceptual map, thematic map, source impact, author relevance, and global citations as well as the most cited countries. The word cloud is a data representation that is regularly utilized for displaying metadata from widely used databases such as Scopus as well as Web of Science [Ahmi, 2022]. It is necessary for perceiving the most frequent terms and organizing them alphabetically [Takawira, 2022]. It shows the author's keywords as well as keywords plus. The latter (keywords-plus) which are automatically computed by a computer system, are those phrases that feature more frequently throughout the names of the manuscript's references rather than the article's title itself. Author Keywords are phrases provided by the original authors that indicate the research topic as well as the research themes [Ahmi, 2022].

The conceptual map was adopted to extract relevant information from data and the representation of that knowledge through intuitive maps or visualizations such as social networks, dendrograms, and bi-dimensional maps [Ahmi, 2022]. A theoretical architecture map was established, including a visual presentation of the contextual framework of all the words that appeared frequently in articles consisting of the spectrum-sensing topic and cognitive radio networks. This was accomplished by employing regional mapping to map the link between two words [Takawira, 2022].

A thematic map consists of four themes defined based on the quadrant within which they are positioned [Takawira, 2022]. The motor themes are those placed in the upper-right quadrant. They are recognized for their high density and centrality [Rusydiana, 2021]. This means they are essential and improved in the field of research [Rusydiana, 2021]. The upper-left quarter themes are classified as highly developed or specialized themes [Rusydiana, 2021]. They have a high density of internal links but few external links, making them of limited importance (low centrality) [Takawira, 2022]. The lower-left quarter themes are classified as declining or emerging [Rusydiana, 2021]. They are underdeveloped and marginal because of their low density and centrality [Della Corte et al., 2019]. All lower-right sector themes are classified as basic themes [Ahmi, 2022]. They are distinguished by their low density and high centrality [Rusydiana, 2021]. Such themes are essential for different research domains because they include all the topics that traverse the various research areas [Shaukat et al., 2020].

## 4. Results

### 4.1. *Source Impact*

This study calculated the h-index of all the journals that spectrum sensing and cognitive radio articles, which is depicted in Fig 7. With an h-index of 9, IEEE Access ranked first in terms of impact. The sensors journal ranked second with an impact factor of 6. This is supported in Fig 8 which indicates the most relevant journal (the journal containing the most relevant articles). Again, it can be noted that IEEE Access is the most relevant source followed by the Sensors journal.

| Element | h_index | g_index | m_index | TC | NP | PY_start |
|---|---|---|---|---|---|---|
| IEEE ACCESS | 9 | 12 | 1.500 | 194 | 21 | 2017 |
| SENSORS | 6 | 9 | 1.000 | 93 | 10 | 2017 |
| IEEE TRANSACTIONS ON VEHICULAR TECHNOLOGY | 3 | 3 | 0.500 | 166 | 3 | 2017 |
| WIRELESS COMMUNICATIONS & MOBILE COMPUTING | 3 | 7 | 0.600 | 55 | 10 | 2018 |
| ENTROPY | 2 | 2 | 0.667 | 7 | 2 | 2020 |
| EURASIP JOURNAL ON WIRELESS COMMUNICATIONS AND NETWORKING | 2 | 2 | 0.667 | 18 | 2 | 2020 |
| IEEE SYSTEMS JOURNAL | 2 | 2 | 0.400 | 31 | 2 | 2018 |
| IEEE TRANSACTIONS ON COMMUNICATIONS | 2 | 2 | 0.333 | 25 | 2 | 2017 |
| SYMMETRY-BASEL | 2 | 2 | 0.333 | 18 | 2 | 2017 |
| 2017 INTERNATIONAL CONFERENCE ON IDENTIFICATION, INFORMATION AND KNOWLEDGE IN THE INTERNET OF THINGS | 1 | 1 | 0.200 | 4 | 1 | 2018 |

*Fig. 7:* Source Impact

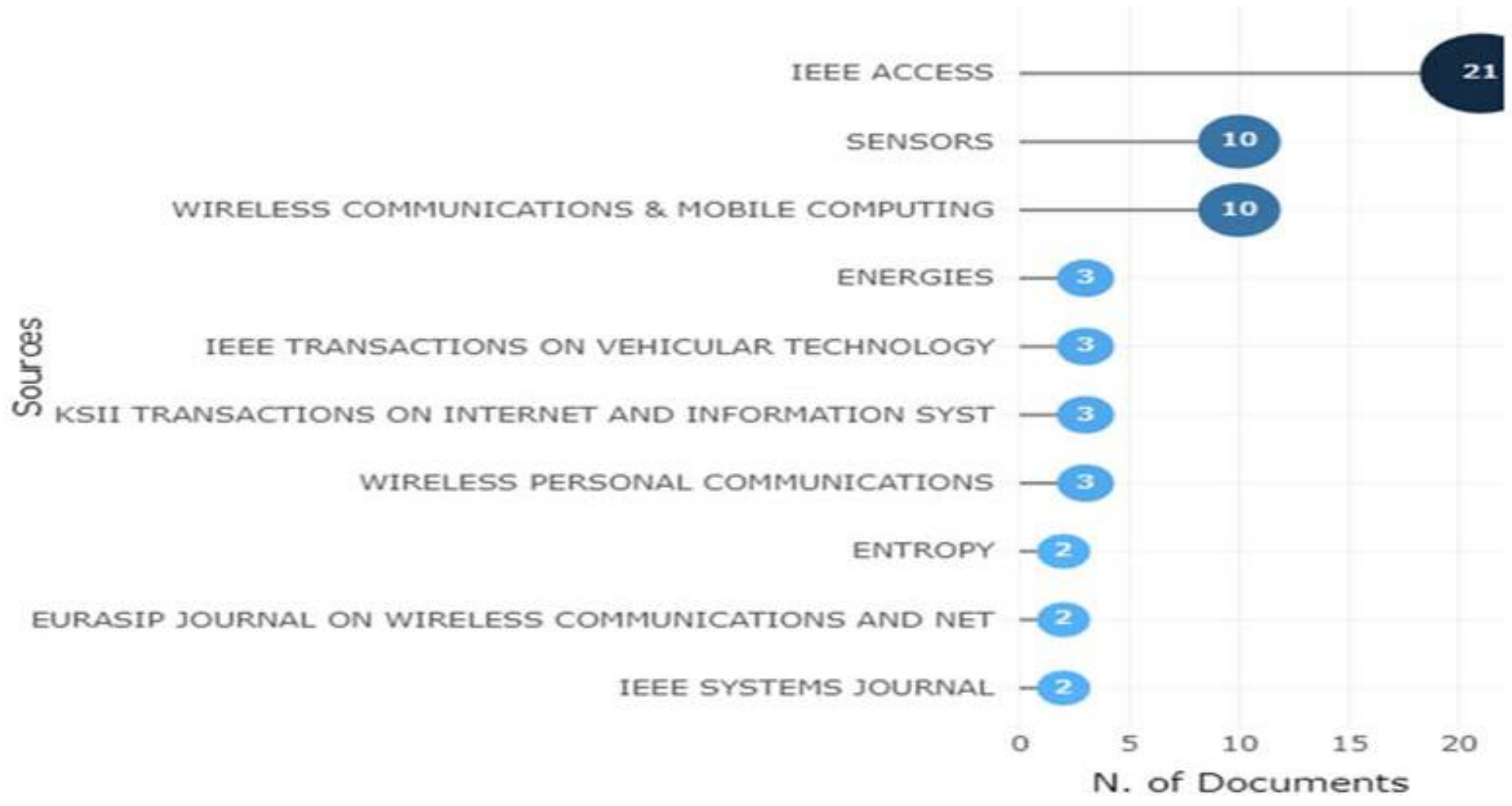

*Fig. 8*: The most relevant sources.

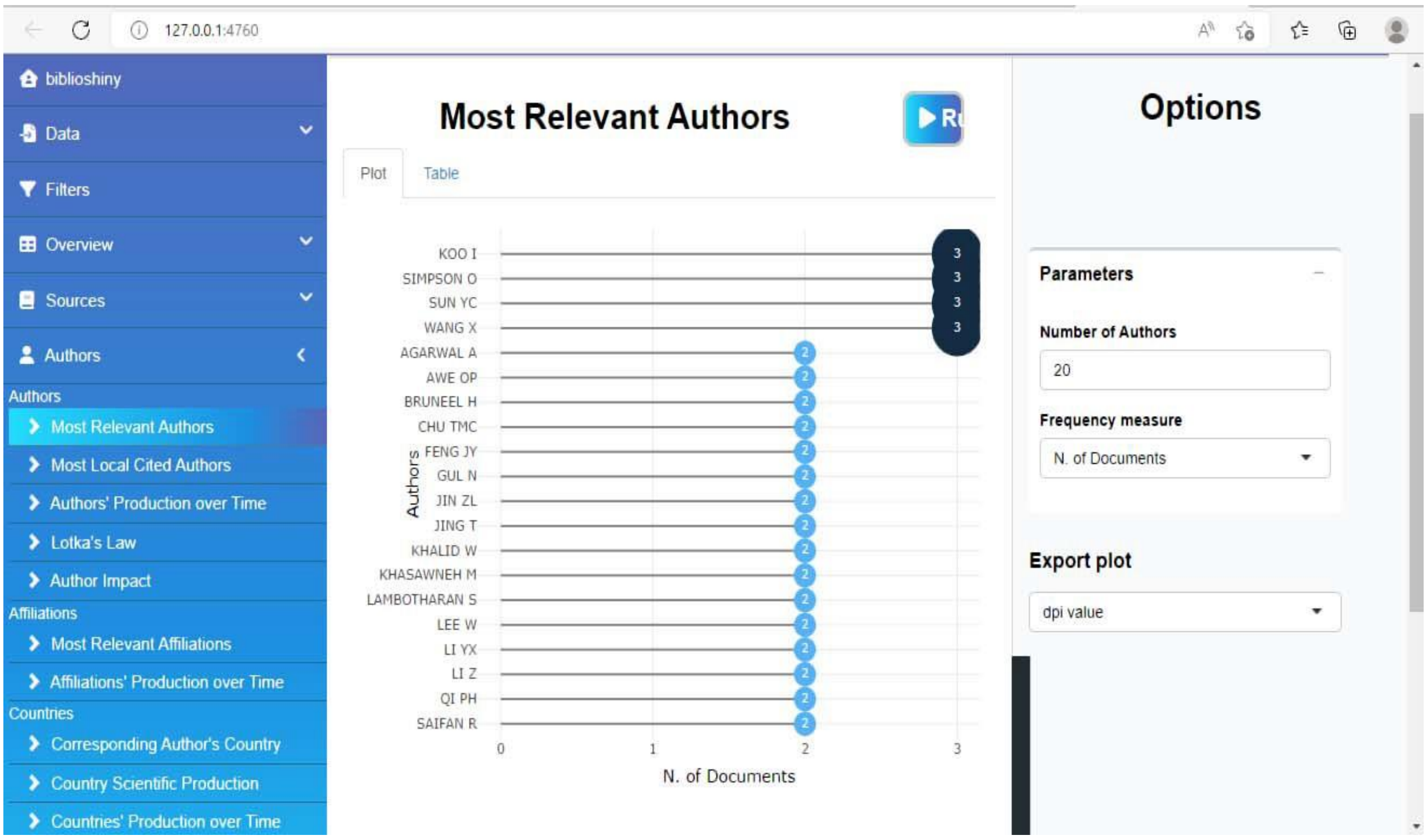

*Fig 9*: The top 20 relevant authors

## 4.2. *Author Relevance*

The top 20 authors were ranked according to their relevance, that is, the number of articles contributing to the body of knowledge (Fig 9). Koo, Simpson, Sun, and Wang proved to outperform other authors in terms of relevance the most, with authorship of 3 articles during the period. This is unsurprising given the authors' obvious strategic advancements in spectrum sensing optimization.

## 4.3. *The most globally cited documents*

The top 20 articles were then ranked by global citations, which is the total of citations by scholars globally (worldwide). In light of this, Fig. 10 reveals that Zhang's article is the most cited, which is 5 times the next-highest ranked work. This demonstrates the author's unique contribution to the body of knowledge.

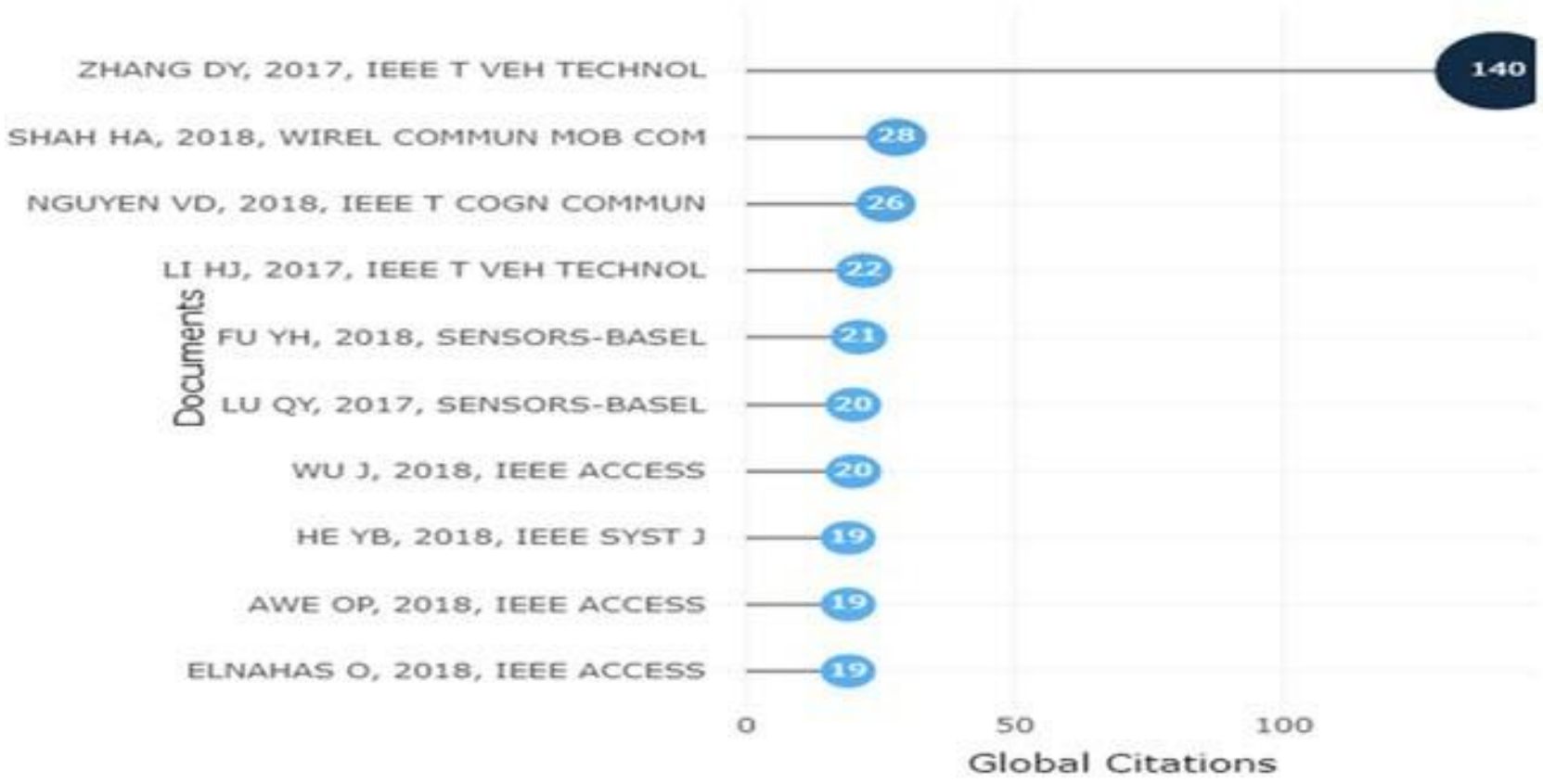

Fig 10: Global citations

## 4.4. *Citations by country*

The top 10 countries were then rated by a total number of citations. Fig. 11 shows that China has the most cited articles, followed by Korea. Interestingly, South Africa is the lowest-rated African country, 5 citations less than its counterpart, Egypt. This demonstrates that Asia has the greatest contribution relative to other continents.

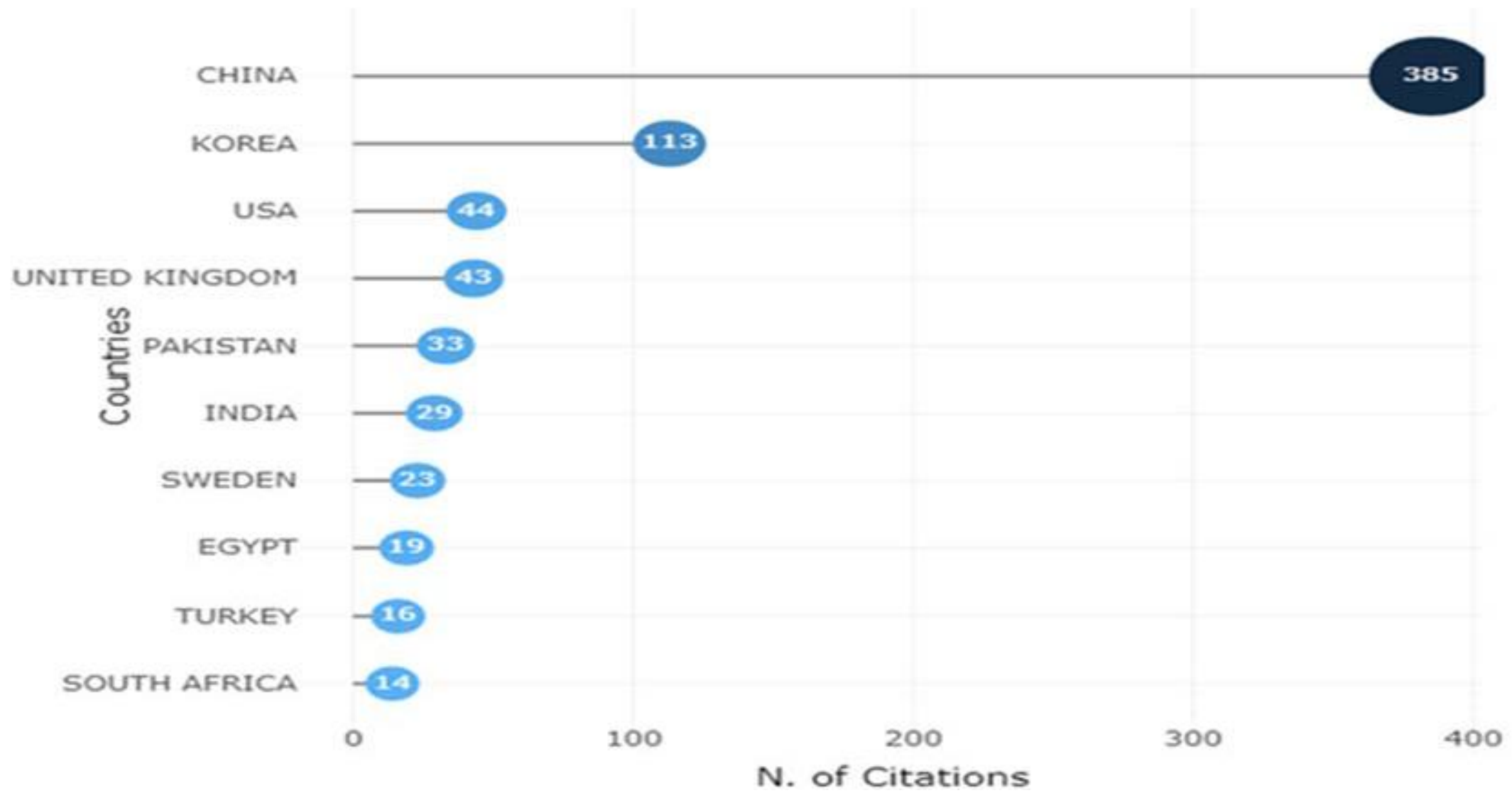

Fig 11: The most cited countries

### 4.5. *Word Cloud*

Fig. 12 indicates a cloud of words that visualizes the most frequently appearing words in the articles on spectrum sensing and cognitive radio networks. The most frequently used word was "access," "optimization," came second "allocation," was third, "algorithms," was fourth, and "energy detection" came fifth. The cloud exposes words in varied sizes dependent on their frequency of appearance. The word placement is arbitrary, however, because of their great size, the dominant terms are placed in the middle for enhanced visibility. Fig. 13 shows the tabular representation of the word cloud.

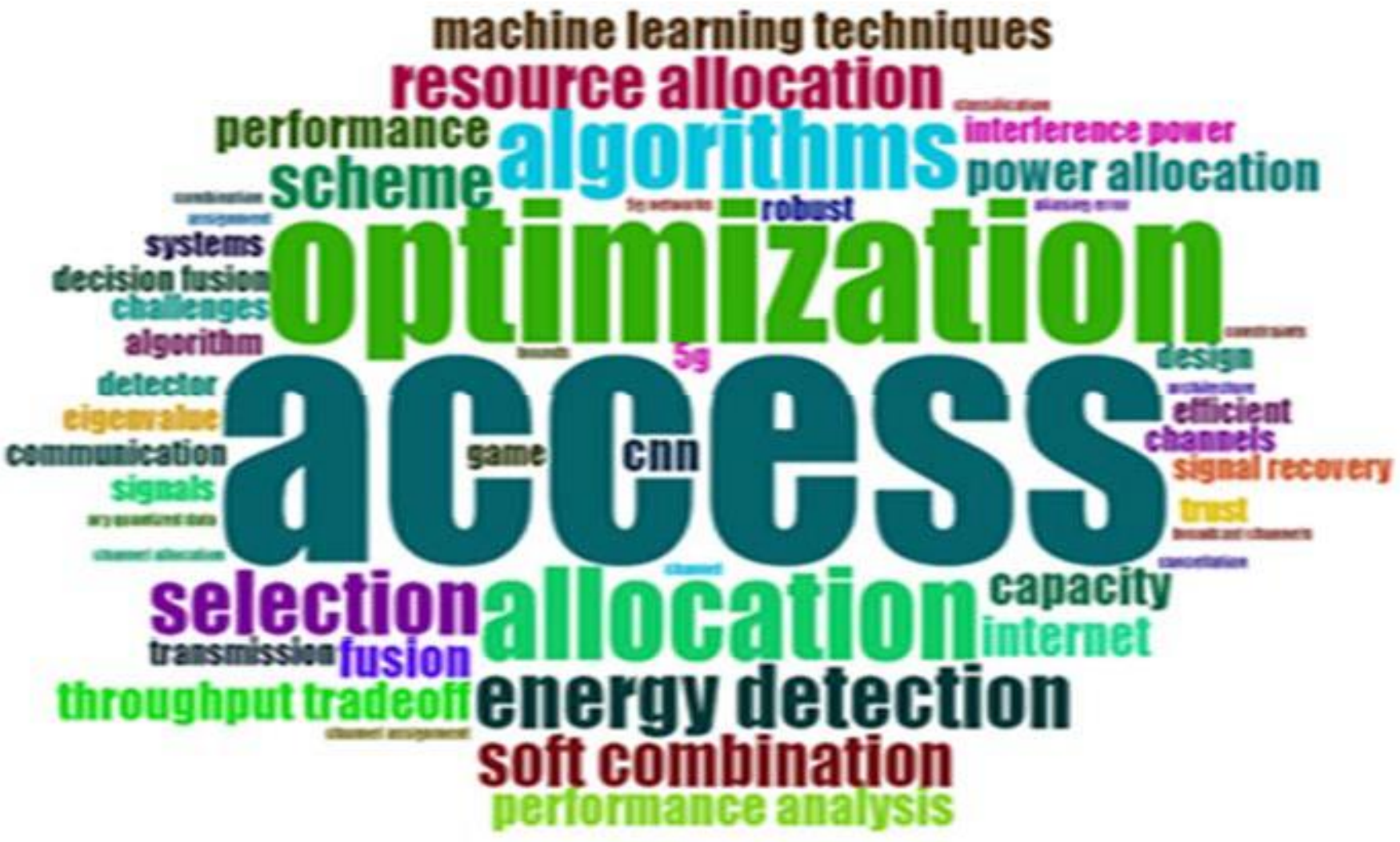

*Fig 12*: showing the most frequently appearing words.

| Terms | Frequency |
|---|---|
| access | 19 |
| optimization | 11 |
| allocation | 7 |
| algorithms | 6 |
| energy detection | 5 |
| selection | 5 |
| resource-allocation | 4 |
| scheme | 4 |
| soft combination | 4 |
| cnn | 3 |

*Fig 13*: word frequency.

### 4.6. *Thematic Theme*

A thematic map based on centrality and density was generated by the biblioshiny tool. An algorithm was applied to present detailed variant information based on examination reference titles and keywords. The specialized (direct) themes under investigation were "capacity" and "signals," as evidenced by the minimal centrality and the large density within the top-left portion. Motor themes for instance "access," " optimization," and "allocation." are located in the upper right region. It is distinguished by its dense population and central location. There are specialized (direct) themes under research indicated by the high density and low centrality in the upper left top region which are "capacity" and "signals ." Themes or topics in the lower left quadrant, such as "eigenvalue" and "machine-learning techniques," are emerging or declining topics. The lower right region contains basic themes such as "energy detection" as indicated by its low density as well as high centrality. Fig. 14 shows the thematic map.

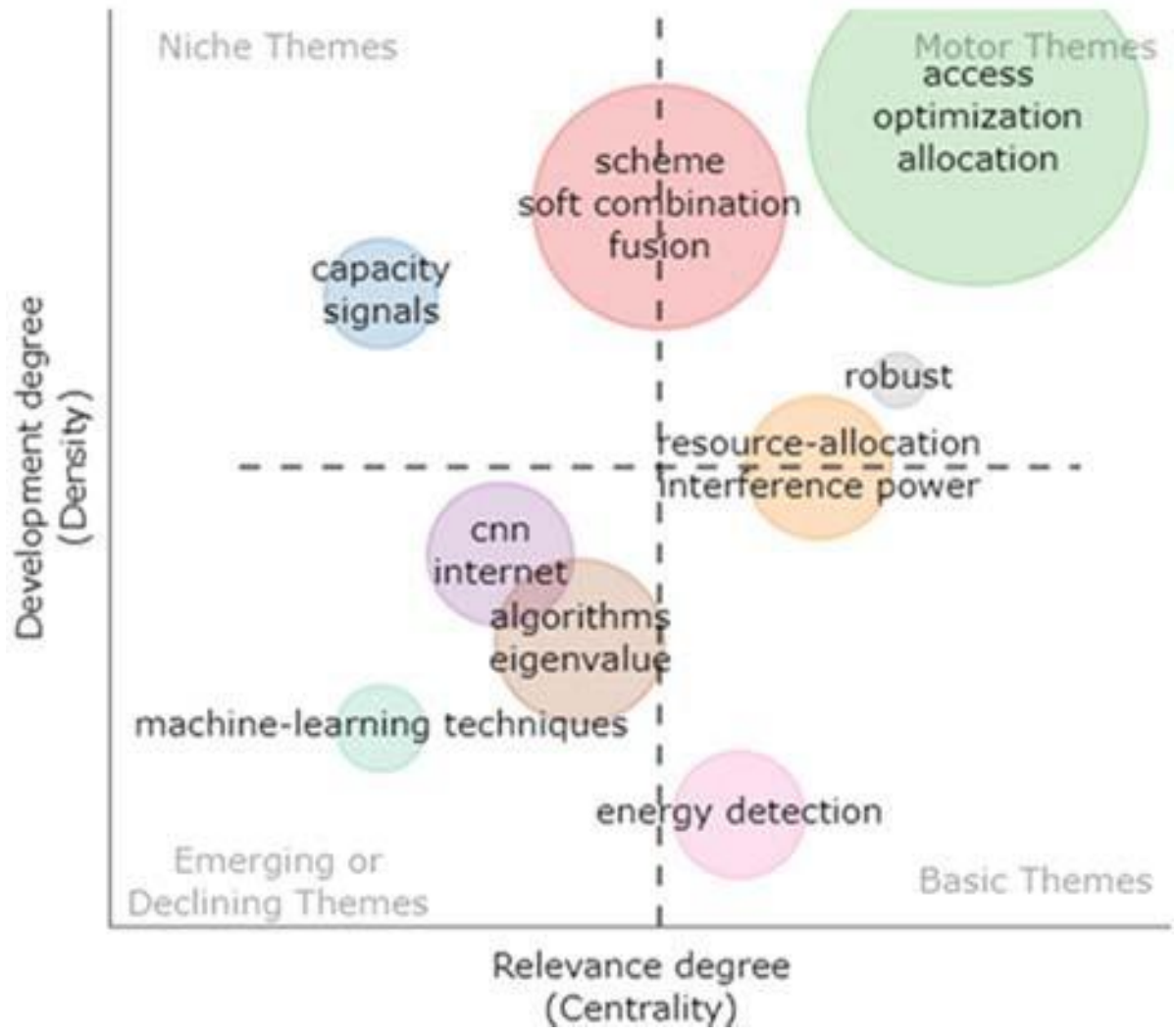

*Fig 14*: Thematic map

### 4.7. *The conceptual map*

A conceptual map shown in Fig. 15 was created using regional mapping to demonstrate the connection between each word that frequently appeared in articles on spectrum sensing in wireless networks. Relevant terms are linked based on their Dim (diminutive) 1 as well as Dim 2 values, where Dim is a small particle, a bibliometric science term, leading to a linkage of words with almost identical values. The red portion consists of various, related words, indicating that many papers conveyed relationships among the words which are listed in this portion. It can be noted that "wireless networks" and "resource allocation" for instance portray a relationship between widely used words.

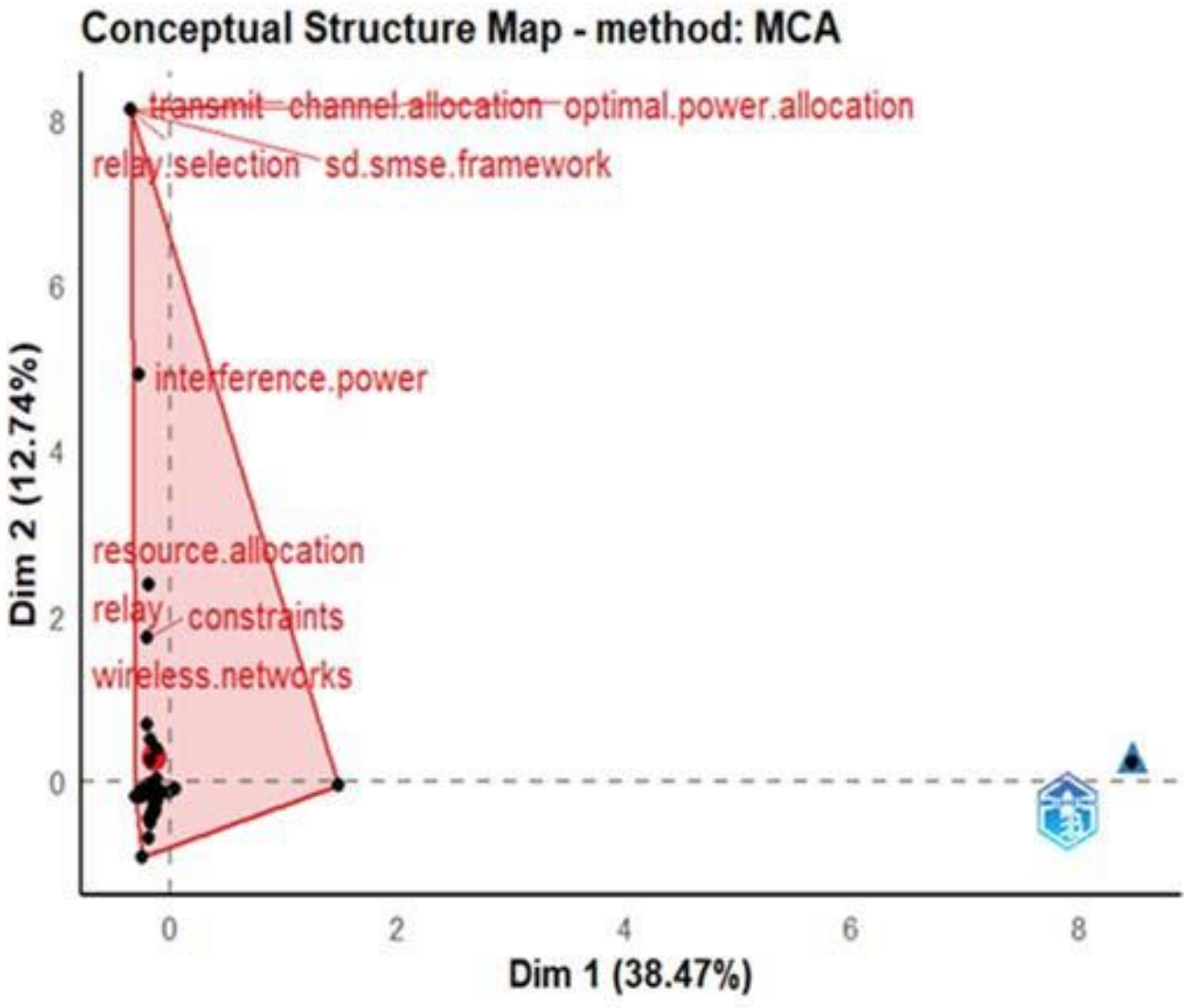

*Fig 15*: Conceptual structure map

## 5. Discussion

The study presents a bibliometric analysis of journal articles on the spectrum sensing notion using the Biblioshiny app. Spectrum sensing has become a prominent focus of researchers' attention due to increased spectrum demand. The literature on spectrum sensing optimization has grown in popularity in recent months. According to the findings, research on spectrum sensing in cognitive radio networks has been widely published by a variety of journals and numerous authors, covering a wide range of specific topics. The IEEE Access journal was found to be the most relevant journal, as it published articles by a large number of authors on the topic. Furthermore, the IEEE Access journal ranked first in terms of impact, having an h-index of 9. This makes it beneficial to scholars seeking references for research on this topic.

The frequently applied phrases were "access," "optimization," "allocation " "algorithms," and "energy detection." Therefore, it can be concluded that these words appeared frequently in the research topics. This implies that the presented machine-learning algorithms can be used for the optimization of spectrum-sensing as compared to conventional methods such as energy and matched filter detectors. Motor themes for instance "access," " optimization," and "allocation" are located in the upper right region. They are distinguished by their dense population and central location. These topics should be further developed due to their importance for future research. This provides a good opportunity for researchers who are looking for unresolved problems in the area of spectrum sensing. The direct / niche themes indicated by the low centrality and high density within the upper left top region are "capacity" and "signals." These under-represented topics are nevertheless areas of rapid development. Themes or topics in the lower left quadrant, such as "eigenvalue" and "machine-learning techniques," are emerging or declining topics. In this case, this is owing to the state-of-the-art machine-learning techniques that have gained popularity in the optimization of spectrum sensing. Understandably, the use of traditional spectrum sensing algorithms such as the eigenvalue is recently declining. This is indicated by the low density and centrality. The lower right region contains basic themes for research such as "energy detection" as indicated by its low density as well as high centrality.

In summary, the findings reveal that at low signal-noise ratios, techniques which are the LSTM/long short-term memory, random forest CNN (Convolutional Neural Network), and transformers can be adopted to optimize spectrum-sensing in wireless networks. Furthermore, it was noted that hybrid models outperform single models for spectrum sensing. This supports the findings by [Koteeshwari and Malarkodi, 2022; Nasser et al., 2021] whose machine-learning or hybrid techniques were applied to optimize spectrum sensing as compared to traditional methods. This implies that for improving spectrum sensing at low SNR, machine learning / deep learning techniques should be adopted.

## 6. Limitations

The study was inferior to data not drawn being from all databases, but rather from only two major ones (Scopus and Web of Science). Other databases such as ScienceDirect and IEEE Xplore could have been applied for the systematic review. Many pieces of information could be missing. This study was limited to the word cloud, conceptual map, thematic map, source impact, author relevance, global citations, and the most cited countries for analysis. Other bibliometrics features for analysis could have been explored such as annual scientific production, co-citation network, collaboration network, and most locally cited authors, just to mention a few. Again, the data was collected until September 2022, so any subsequent changes were not highlighted in this article. Suggestions for future research include using multiple databases (sources) for bibliometric analysis. Other keywords about spectrum sensing or the synonyms for the keywords used in this study could have been used. This study was restricted to articles written in English. To avoid omitting articles that would have been beneficial for analysis, an attempt to translate the non-English articles could have been implemented.

## 7. Conclusion

Several authors propose machine-learning algorithms for optimizing spectrum sensing, according to this study. Hence these keywords were commonly employed in research topics and have an opportunity for further development. The number of articles on the theme of spectrum sensing published by journals is substantial and has the potential to develop further, considering the ongoing improvement of traditional techniques.That being said, machine learning techniques / hybrid models were found to optimize spectrum-sensing in wireless networks. Therefore this study recommends researchers from diverse continents contribute to the notion of the improvement of spectrum sensing. A thorough review of the literature should be implemented to identify the drawbacks of the existing spectrum sensing techniques. Various databases and search criteria should be utilized to gather a wide spectrum of articles for review. Remarkably, this study reveals that traditional spectrum-sensing approaches have several limitations, for instance, poor performance at low SNR. In summary, energy detection performs poorly when the SNR ratio is low, cyclostationary features and wavelet-based are complex, and matched filtering necessitates some licensed user signal information prior to sensing. Motivated by these drawbacks, researchers should focus on implementing machine-learning or hybrid models for the classification of wireless spectrum. Researchers should also focus on real-time, cooperative spectrum sensing in 5G and 6G wireless networks.

## Acknowledgments

This study was carried out at the University of Fort Hare, Department of Computer Science and was funded by the ACTS AI4D and the NRF SA. The researcher owns the deductions, ideas, and findings presented in this research and accepts full responsibility.

## Conflicts of interest

The authors have no conflicts of interest to declare.

## Authors Profile

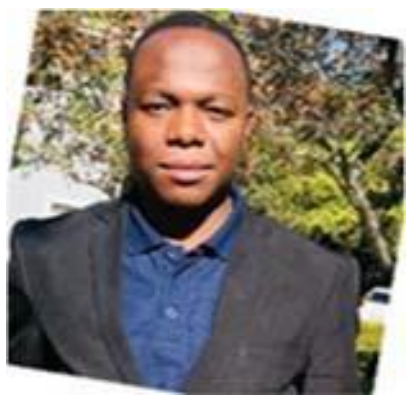

**Nyashadzashe Tamuka** is a final year Ph.D. in Computer Science student at the University of Fort Hare, South Africa. He obtained his BSc. Honours in Computer Science and MSc in Computer Science at the University of Fort Hare. He possesses 5 years of research and teaching experience. His research interests are in Machine Learning and wireless networks.

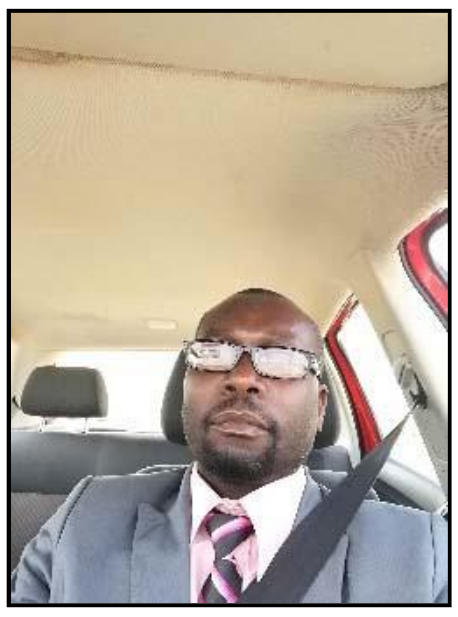

**Prof. K Sibanda** is a well-experienced Walter Sisulu University academic, whose research interests are in Artificial Intelligence with a bias to machine learning. In his career, he has supervised to completion 5 PhD and 19 MSc in Computer Science students. Most of his research centers around machine learning techniques for classification, clustering and prediction. He has begun exploring such techniques for educational solutions. He believes that machine learning approaches are very robust in answering a wide range of problems including teaching and learning problems.